\documentclass[twocolumn, noshowpacs,showkeys, amsmath,amssymb]{revtex4}
\newcounter{qcounter}
\newcommand{\bra}[1]{\langle #1|}
\newcommand{\ket}[1]{|#1\rangle}
\newcommand{\braket}[2]{\langle #1|#2\rangle}

\usepackage{graphicx}
\usepackage{dcolumn}
\usepackage{bm}
\usepackage{url}
\usepackage{natbib}
\usepackage{amsmath}
\usepackage{amssymb}
\begin{document}
\begin{abstract}
Quantum mechanics under the Copenhagen interpretation is one of the most experimentally well verified formalisms. However, it is known that the interpretation makes explicit reference to external observation or ``measurement." One says that the Copenhagen interpretation suffers from the \emph{measurement problem}. This deficiency of the interpretation excludes it as a viable fundamental formalism and prevents the use of standard quantum mechanics in discussions of quantum cosmology. Numerous alternative interpretations have been developed with the goals of reproducing its predictive success while obviating the measurement problem. While several interpretations make distinct, falsifiable, predictions, many claim to precisely reproduce the results of standard quantum mechanics. The sheer number of interpretations raises several issues. If the experimental predictions are identical, how are they to be assessed? On what grounds can an interpretation be said to trump another? Without recourse to experimental findings, one may continue to assess an interpretation on its logical structure, self-consistency, and simplicity (number and plausibility of its assumptions). We argue, and where possible, demonstrate, that \emph{all} common interpretations have unresolved deficiencies. Among these deficiencies are failures to resolve the measurement problem, fine-tuning problems, logical/mathematical inconsistencies, disagreement with experiment, and others. Shortcomings as severe as these call into question the viability of any of the common interpretations. When appropriate, we indicate where future work may resolve some of these issues.   
\end{abstract}

\title{Why Current Interpretations of Quantum Mechanics are Deficient}         
\author{Elliott Tammaro} 
\email{tammaroe@chc.edu}
\affiliation{Department of Physics, Chestnut Hill College, Philadelphia, Pennsylvania, USA}   
 \date{Received: date / Accepted: date}
\pacs{.23}
\maketitle


\section{Introduction}
At this point in time it appears that a stalemate has been reached with regard to the interpretation of quantum mechanics. Surprisingly, despite the roughly ninety years since its conception, there is currently no single widely accepted interpretation. The variety of interpretations has acted to divide the physics community into camps. For example, one might be a ``Bohmian" or an ``Everettian" or in the ``I shut up and calculate" camp. There is virtually no travel between camps, but there is much in the way of campaigning for new recruits. In addition to being a mere inconvenience, we currently stand at the cusp of physics beyond the standard model and it may be that further advancement will demand a deeper understanding of $20^{\mathrm{th}}$ century physics. It is firmly established that string theory, while still the most promising attempt at unification, does not provide any deeper insight into quantum mechanics. While we recognize that string theory might play an essential role in physics in the $21^{\mathrm{st}}$ century, there are indicators that a deeper understanding, or more ideally, a complete resolution, of the measurement problem may be a requirement for future theories. Included in areas where further insight into the measurement issue is needed are approaches to quantum cosmology and quantum gravity.     

We proceed by asking an obvious question, is there a single correct interpretation of quantum mechanics? Or may one be a Bohmian on Monday, Wednesday, and Friday, an Everettian on Tuesday, Thursday, and Saturday and side with consistent histories on Sunday? On what grounds may an interpretation be said to trump another interpretation? There are two methods by which one may assess an interpretation. Firstly, several interpretations (more appropriately new theories) make predictions that are distinct from those of quantum mechanics under the Copenhagen interpretation, which makes falsification by comparison with experiment straightforward. We will highlight examples of such when they occur. Ideally, it is hoped that each interpretation will be found to make different experimentally falsifiable claims. Presently, however, many of the interpretations reproduce the results expected from standard quantum mechanics. If consistency with experiment at any cost is what is valued, then several interpretations survive. Fortunately, there is a longstanding history in physics which dictates that without direct experimental evidence an appeal to logic, consistency, or simplicity may be used to whittle the playing field of theories lacking such qualities. Of particular note for this work are problems with fine-tuning, wherein, generically speaking, a theory requires a highly specialized initial state, and questions of consistency of the mathematical and logical structure. It is found upon examination that \emph{no current interpretation} is consistent with experiment, resolves the measurement problem, and is completely free from logical deficiencies or fine-tuning problems.

Discussions promoting one interpretation over others appear in many sources. Too often the authors of such works are highly biased in favoring one interpretation over another. Our aim in this note is to provide an objective examination of the most prevalent interpretations with the goals of highlighting implicit assumptions, unveiling inherent deficiencies, and indicating where future work may resolve such issues. We mention in passing that this present work is not intended to review arguments which appear elsewhere in the literature, although minor overlaps do occur. We begin in section \ref{CI} with the Copenhagen interpretation, which is not only a commonly held interpretation, it is also the interpretation that is taught to each student embarking on the study of quantum mechanics. 

\section{Copenhagen Interpretation/ Relational Quantum Mechanics/ Ensemble Interpretation} 
\subsection{Copenhagen interpretation}
\label{CI}
The usefulness of the Copenhagen interpretation is undeniable. It allows one to make probabilistic predictions for the results of measurements, given the wavefunction for a system, via the Born rule. However, it is evident that the reference to external observation in the form of ``measurement" raises questions about the universal validity of the interpretation. The collection of difficulties arising from the inclusion of measurement as fundamental is known as the \emph{measurement problem}. Before we make additional comments, it is helpful to review quantum mechanics under the Copenhagen interpretation.

Quantum mechanics under the Copenhagen interpretation purports the following. Isolated physical systems are completely described by a Hilbert space element (a ray in a Hilbert space, to be precise) $\ket{\psi}$, referred to as the ``wavefunction." The wavefunction evolves in time according to two distinct processes. We will match the literature and refer to them as \emph{Process 1} and \emph{Process 2}\cite{HughEverett1957}. \begin{list}{\emph{Process \arabic{qcounter}}.~}{\usecounter{qcounter}}\item Discontinuous, indeterministic time evolution which sends $\ket{\psi}$ into an eigenstate $\ket{o_i}$ of observable $O$ with probability $|\braket{o_i}{\psi}|^2$ as a result of a measurement of $O$.\item Unitary time evolution --- the time dependence of $\ket{\psi(t)}$ is governed by the Schr\"odinger equation, $i\hbar\dot{\ket{\psi(t)}}=H\ket{\psi(t)}.$\end{list} 

Let us now demonstrate that these two processes, as stated, are inconsistent. Consider an isolated observer $A$, initially in state $\ket{A_{in}}$ and system of interest $S$, initially in state $\ket{S_{in}}$. Whereby the term ``observer" refers to any device capable of inducing \emph{Process 1}. Since $A$ induces \emph{Process 1} the result of interaction between $A$ and $S$ must be a mixed state \begin{equation}\rho=\sum_i|\braket{o_i}{S_{in}}|^2\ket{A_{f} (i)}\bra{A_{f}(i)}\otimes \ket{o_i}\bra{o_i}\label{1}\end{equation} where we have assumed that $A$ observed $O$ and we have indicated the final state of $A$ with $\ket{A_{f}(i)}$. If, on the other hand, \emph{Process 2} is valid for all isolated systems then it \emph{must} too be valid for the assumed isolated system consisting of $A+S$, thereby dictating that the final state is \begin{equation}U_{AS}\ket{A_{in}}\otimes\ket{S_{in}}\label{2}=\sum_j\ket{o_j}\ket{a_j}\end{equation} where $U_{AS}$ is the $A+S$ system time evolution operator and $\ket{a_j}$ corresponds to the apparatus reading eigenvalue ``$o_j$". The process indicated in equation \eqref{2} is referred to as a Von Neumann measurement \emph{of the first kind}. What is the state of the $A+S$ system after the measurement? It is clear that possible final states \eqref{1} and \eqref{2} are mutually exclusive (if a system is in state \eqref{1} then it is not in state \eqref{2} and vice versa.) because the final state generated by \emph{Process 1} is mixed, while that generated by \emph{Process 2} is pure. It is well known that no unitary evolution can transform a pure state into a mixed state. We demonstrate that a logical inconsistency arises because of incompleteness. \emph{Processes} \emph{1} and \emph{2} result in \emph{physically} distinct states. According to Process \emph{2} the state of the system of interest remains in superposition. As a consequence, the possibility of interference exists. If Process \emph{1} occurs then the system of interest, in conjunction with the apparatus, is no longer in superposition and the possibility of interference is eliminated. There is nothing inherent to the theory that either rejects the mutual application of \emph{Processes 1} and \emph{2} or selects one over the other. Since the Copenhagen interpretation fails to make a firm prediction about the final state, it is appropriately deemed incomplete. Although performing an experiment that determines the physically realized final state of the $A+S$ system is current an impossibility, eventually it will be possible. We will determine which process, \emph{1} or \emph{2}, or some hitherto unconsidered possibility, is actually realized. It is guaranteed that in this way the Copenhagen interpretation will be found incomplete in a laboratory setting.

Additional deficiencies arise from attempts at reconciling \emph{Process 1} and \emph{2}. Consider a second observer, observer $B$, who will function as a ``Wigner's friend" for the $A+S$ system. We assume that observer $B$ is equivalent to observer $A$ \emph{in all respects}. At time $t_{in}$ observer $A$ interacts with the system of interest $S$, thereby making a measurement. Assume without loss of generality that observer $A$ found definite value $o_i$ \emph{at time $t_{in}$}. At time $t_{f}$ ($t_f>t_{in}$) let observer $B$ interact with the $A+S$ system in an attempt to determine which value observer $A$ determined for $S$. As a ``Wigner's friend," assume, without loss of generality, that observer $B$ discovers that the system $S$ is in state $\ket{o_i}$. The question arises, what time should observer $B$ assign to the observation that $S$ is in $\ket{o_i}$? If unitary dynamics holds good then, according to $B$, system $S$ was not in a definite state $\ket{o_i}$ \emph{until} $t_{f}$, in contradiction to the claim of $A$ that it was in $\ket{o_i}$ at $t_{in}$. One is either forced to abandon a unitary description of observer $A$ from the standpoint of observer $B$ or one is forced to abandon the equivalence of observers $A$ and $B$. This presents difficulties especially in the context of a relativistic theory. In relativistic theories, the finite speed of light demands that (local) observers make local observations, in order to collect information about distant events \cite{Einstein}. We find that consistency with this premise of relativity (that \emph{equivalent} observers make local observations) enforces the abandonment of a unitary description of the observer.    
   
We refer to the set of inconsistencies that arise from the attempted simultaneous application of \emph{Processes 1} and \emph{2} as the \emph{P1}\&\emph{2 problem}.

\subsection{Relational Quantum Mechanics}
The standpoint of relational quantum mechanics is that both \emph{Process 1} and \emph{Process 2} occur and are correct because the state of a system is not an absolute quantity, but may only be given in reference to a particular observer, analogously to velocity \cite{Rovelli}. That the state of a system is a ``relative" quantity is not in question for our current purposes (although we note in passing that for the relational view to be maintained the quantum state equivalent of a Lorentz transformation, which would permit the wavefunction \emph{with} respect to observer $A$ to be transformed into the wavefunction with respect to observer $B$, must exist and no such quantity has been purported). Instead we concern ourselves with the fact that relational quantum mechanics demands that \emph{1} and \emph{2} are simultaneously correct. As we have indicated, these two processes yield different experimental results. Thus, to maintain them both simultaneously is a logical fallacy (assuming two mutually exclusive propositions to be true). Since experiment will eventually reveal which process trumps the other (or that some other process occurs), we see that this view will be falsified. Bluntly, relational quantum mechanics suffers from the \emph{P1}\&\emph{2} problem (see Section \ref{CI}). Since it makes no attempt to clarify the measurement process nor provides an alternative, it fails to resolve the measurement problem. 

\subsection{Ensemble (Statistical) Interpretation}
The ensemble or statistical interpretation is often cited as a minimalist interpretation \cite{Ballentine}. It states that the wavefunction does not apply to individual systems, but only to an ensemble $E$ of similarly prepared systems. The hope is that one may resolve the measurement problem, at least with regard to collapse, by imagining that the individual subsystems always have well defined properties, but that by some as-of-yet unknown mechanism they are statistically correlated in the manner prescribed by the Born rule \cite{Ballentine}. An immediate deficiency may be noted. Recall that Gleason's theorem demonstrates that any interpretation capable of reproducing the results of quantum mechanics under the Copenhagen interpretation must be \emph{contextual} \cite{Gleason}. That is, the result of a measurement must depend on the measuring device used to measure it. If the subsystems have properties defined \emph{without} reference to a measuring device, then one has introduced \emph{noncontextuality} and it is guaranteed, via Gleason's theorem, that the interpretation disagrees with quantum mechanics in an experimentally falsifiable manner. If, in order to avoid this aforementioned deficiency, contextuality is introduced, then one may charge this interpretation with being incomplete. Namely, if contextuality is introduced, then systems post measurement are different than the systems immediately before the measurement occurred, but the ensemble approach offers no description of this (collapse-type) process. Requiring a collapse-type process, but without an attempt to reduce it to either purely unitary or possibly fundamental non-unitary processes, indicates a failure to resolve the measurement problem. Note that on aesthetic grounds this approach is dissatisfying because it offers no explanation for or description of the \emph{mechanism} of statistical correlation among the subsystems.   
   
\subsection{Quantum Bayesianism}
Quantum Bayesianism, or QBism, concerns the application of Bayesian probability (statistical inference/updating) to quantum mechanics. In particular, an agent (read gambler) acts as a classical observer embedded in an otherwise quantum world. The state vector represents the degree of belief that an agent has for events to occur upon measurement, so that wavefunction collapse is understood as a jump in the knowledge of the observer. It is an essential point that no nonlocal behavior exists, including nonlocal correlation \cite{Fuchs1}\cite{Fuchs2}\cite{Fuchs3}. 

QBism shares so much in common with the Copenhagen interpretation that it cannot rightfully be called a distinct interpretation. In particular, it uses a notion of measurement that corresponds precisely to that of the Copenhagen interpretation. No refinement in understanding of the measurement process is introduced. That is, there is no attempt at describing measurement in terms of more fundamental processes. Just as in the Copenhagen interpretation, the Born rule is assumed without justification. The Copenhagen interpretation is notable for a (decidedly implausible) sharp cut between quantum and classical worlds, and such a sharp divide indeed remains in QBism. The agents of QBism function as a classical observers. As such, the sharp divide between the classical and the quantum occurs ``at the agent." That is, one \emph{cannot} apply quantum mechanics to describe the function of the agents. This is an odd and unappealing conclusion as it is certainly believed that physical agents are constructed from protons and electrons interacting in a manner that is not different from other matter and yet QBism forces us to refrain from applying quantum mechanics to the agents because of their status. It is straightforward to see that the commonalities between QBism and the Copenhagen interpretation imply that they share deficiencies. In particular, QBism suffers from the \emph{P1}\&\emph{2} problem.

A fundamental premise of QBism is the change in the definition of probability from the frequentist inference (probability is defined as the ratio of the number of elements with property $X$ to the total number of elements) to the Bayesian inference (probability is defined as a degree of belief in a given statement). As a result, the collapse of the wavefunction is assumed \emph{nonphysical}. We find this view untenable. The wavefunction after collapse represents a radically different physical system than before collapse. Consider a gambler betting on a horse race. Assume she has some (incomplete) data on each horse. Her bets are distributed according to the data. If she is given new information about the horses, her bets will generically be different. Such is the case with wavefunction collapse in QBism. However, the gambler's bets have no effect on the outcomes of the races, and as such the analogy breaks down. Finally, the claim that the collapse is a result of the changing knowledge of the observer (agent) contradicts the well verified dictum that \emph{knowing the wavefunction of a system represents a state of complete knowledge of system}. In standard quantum mechanics and quantum statistical mechanics, states of incomplete knowledge (epistemic information) are represented by density matrices (mixed states). Contrary to this extremely successful model, QBism claims that pure states \emph{also} represent incomplete knowledge. It is apparent that these definitions are mutually exclusive, and it is unclear how the distinction is made manifest in QBism.       

\section{Decoherence}
\label{Decoherence}
The decoherence program attempts to bypass the difficulties associated with the Copenhagen interpretation by claiming that while the evolution of a perfectly isolated system is governed by Process \emph{2} alone, no quantum system is truly isolated during the measurement process. It is, instead, an open system actively interacting with a quantum environment. Environmental perturbation destroys coherence among superposed states by spreading the initial coherence throughout the environment, whose state is inaccessible experimentally. It produces mixed states when the environmental degrees of freedom are traced out \cite{ZurekI} \cite{ZurekII} \cite{ZurekIII}. More explicitly, consider, as before, a system $S$ whose initial state is $\ket{S_{in}}$, and a measuring apparatus $A$ with initial state $\ket{A_{in}}$. The apparatus is assumed to perform the following function \begin{equation}U_{SA}\ket{S_{in}}\otimes\ket{A_{in}}=\sum_{i}k_i\ket{o_i}\ket{a_i}\label{3''}\end{equation} Where $\ket{o_i}$ is an eigenstate of observable $O$ with eigenvalue $o_i$, $\ket{a_i}$ is the pointer state corresponding to eigenvalue $o_i$, and we have assumed that $\ket{S_{in}}=\sum_ik_i\ket{o_i}$. This Von Neumann measurement of the first kind seems like a promising solution to the measurement problem because it has introduced, via a unitary interaction alone, perfect correlation between the system and the pointer states. However, equation \eqref{3''}, as it stands, suffers from two difficulties. Firstly, equation \eqref{3''} is a \emph{superposition}. Interaction with the measuring apparatus has failed to select a particular value for the observable, \emph{and} interference may still be observed. This may not, at first, seem so problematic because it is often argued that it would be too difficult to observe interference among the superposed states because the apparatus is macroscopic. According to this argument, equation \eqref{3''} is effectively \begin{equation}\rho_S=\sum_i|k_i|^2\ket{o_i}\bra{o_i},\end{equation} which is a mixed state with probabilities dictated by the Born rule. For current technologies it would be very difficult indeed to observe interference with a (near) macroscopic device. Fundamentally speaking, however, equation \eqref{3''} \emph{does} permit the observation of interference. Hence we cannot disregard the superposition in a fundamental description, as future technologies may bring it within experimental grasp. We refer to this deficiency as the \emph{superposition problem}. Secondly, \eqref{3''} suffers from the so-called \emph{preferred basis problem} \cite{Stapp}\cite{Galvan}. We briefly state it here. Since nothing at this point has selected a preferred set of basis states of the system $S$ or of the apparatus $A$, one might imagine re-expressing \eqref{3''} using $\ket{o_i}=\sum_jf_{ij}\ket{o'_j}$ and $\sum_ik_if_{ij}\ket{a_{i}}\equiv k'_j\ket{a'_j}$ (no sum on $j$) as \begin{equation}= \sum_ik'_i \ket{o_i'}\ket{a_i'}\end{equation} where $\ket{o_i'}$ is the $i^{\mathrm{th}}$ eigenstate of a new observable $O'$, and $\ket{a'_i}$ is the pointer state corresponding to $o'_i$. The preferred basis problem is quite serious. It implies that the unitary evolution of quantum mechanics is incapable of indicating if a particular observable was ``observed." This stands in strong contradiction to common experimental procedure. If an experimentalist aims to observe the position of a particle, for example, then a certain device (a position measuring apparatus) is employed and a position measurement follows. That is, it seems that during the measurement process a base \emph{is} indeed selected. The preferred basis problem indicates that equation \eqref{3''} has not only failed to select a particular eigenvalue, it has even failed to select a particular observable! For these reasons \eqref{3''} cannot be referred to as a ``measurement" in any strict sense. 

The decoherence program proposes the following resolution. Let the system and apparatus together interact with an environment $E$ whose initial state is $\ket{\epsilon_{in}}$. The interaction with the environment is unitary and it is assumed that it produces a state \begin{equation}U_{SAE}\ket{S_{in}}\otimes\ket{A_{in}}\otimes\ket{\epsilon_{in}}=\sum_ik_i\ket{o_i}\ket{a_i}\ket{\epsilon_i}\label{3}\end{equation} If such an interaction occurs, then there is only one base of system and apparatus which maintains perfect correlation even after interacting with the environment. Said basis is deemed ``robust" against environmental interaction and is referred to as the \emph{environmentally selected pointer basis}. It is in this fashion that the decoherence program claims to solve the preferred basis problem \cite{ZurekI} \cite{Schlosshauer1} \cite{Schlosshauer2}.  

Does decoherence actually solve the preferred basis problem? We will argue that decoherence does not solve preferred basis problem, and more generally, that unitary dynamics alone cannot do so \cite{Galvan}. First consider a general argument. Quantum mechanics postulates that each system is fully described by a vector (more properly Hilbert space ray) in a Hilbert space, and evolves via unitary action. That ``a system is fully described by a Hilbert space vector" ensures that all systems that are described by the same state vector are \emph{physically} equivalent. If this were not so, then there must be a physical quantity that acts to distinguish between identical state vectors, which is the definition of a hidden variable. As a result, if one is unwilling to accept hidden variables, then basis independence must also be assumed. Furthermore, the unitary sector of quantum mechanics (unitary evolution without collapse mechanism) completely lacks a mechanism to single out a base. That is, there is no type of interaction that might select a basis from a dynamical origin. One is compelled to concede that despite claims otherwise unitary theories do not have preferred bases. Let us argue further. One may tie the freedom to choose a Hilbert space base to the freedom of coordinate system choice. Thusly, that the choice of coordinate system is nonphysical dictates that basis choice must also be nonphysical. The claim that the inclusion of an environment which interacts with the $S+A$ (system and apparatus) system ``selects" a basis is faulty. The full system ($S+A+E$) remains a unitary quantum system and without an additional mechanism to endow different bases with physical significance, all basis choices are valid. 

Given that no mechanism in the unitary sector of quantum mechanics selects a base, then in what sense is the base which allows for the decomposition \eqref{3} ``preferred" to other bases?  The answer put forward in the decoherence program is that the decomposition \eqref{3} is preferred because it has a particularly appealing interpretation from the viewpoint of a human observer. Namely, the measuring apparatus has ``read" the system and is not further disturbed (altered) by the environment. This is both highly ad hoc and circular logic as the base chosen for decomposition within the interpretation is preferred only by the fact that it \emph{permits an interpretation}. Adherents to the decoherence program are asking one to augment unitary quantum mechanic with a non-quantitative rule for base selection that is constructed on subjective grounds. 

One does not know, a priori, the state of system. It is this lack of knowledge that provokes one into ``making measurements" in the first place. Experimentalists must, therefore, ``trust" the reading on the pointer of their apparatus. Can we know if the observed pointer reading accurately represents the ``system value" or was it perturbed (perhaps even strongly) by the environment? Clearly, one cannot know without an appeal to another measuring apparatus, for which the same difficulty may arise. It is universally agreed that only \emph{testable} statements should be included as part of a theory. Can the claim of perfect correlation among system, apparatus, and environment be tested? Since the interpretation rests on this statement (decoherence fails to be an interpretation if this statement does not hold good), then any tests of this statement, which must proceed by not assuming perfect correlation, could not be interpreted. They would be non-interpretable. Consequently, the statement of perfect correlation must be taken as an untestable fundamental postulate, which, within the context of physical theories, is highly dissatisfying and strongly suspect.        

A key point must be highlighted here, \emph{we do not, a priori, know the form of the unitary interaction in \eqref{3}}, yet \eqref{3} supposes that the environment ``Von Neumann measures" the system and apparatus. We wish to emphasize how unlikely it is that this assumption is realized naturally. The measuring apparatus $A$ is \emph{designed} via the interaction exhibited in \ref{3''} to produce a perfectly correlated state. It is obvious to every experimentalist that there is a great difficulty in building a device which acts on a system to produce perfect (or even near perfect) correlation between states without introducing uncontrollable phase shifts. Yet \eqref{3} states that the environment performs this function (idealized Von Neumann Measurement) without trouble. While one might argue that this is possible in principle, it is also highly contrived. What prevents, for example, in a number of experiments, the result \begin{equation}U_{SAE}\ket{S_{in}}\otimes\ket{A_{in}}\otimes\ket{\epsilon_{in}}=\sum_i\kappa_i\ket{o_i}\ket{a_i}\ket{\epsilon_i}\label{3'''}\end{equation} (where $\kappa_i\neq k_i$, and where we still explicitly assume that $\ket{S_{in}}=\sum_ik_i\ket{o_i}$)? Clearly, states such as \eqref{3'''} do not follow the Born rule, and so if they were to occur, decoherence would immediately be at odds with experiment. The decoherence program has the burden to justify \eqref{3}, which is tantamount to deriving the Born rule. What level of justification would prove sufficient? To justify the unitary action in \eqref{3}, adherents to the decoherence program must model the environment as a system of particles which interacts with $S$ in known ways (electromagnetically, or perhaps gravitationally) and from this model deduce that the aforementioned unitary action is possible \emph{and} that it occurs vastly more frequently than any other unitary transformation (so as to remain in agreement with the Born rule). Let us highlight just how unnatural the assumption of \eqref{3} is with an accurate analogy. Let the system of interest and apparatus be considered a boat immersed in the environment, here conceived of as the ocean. Then \eqref{3} indicates that size and shape of the swells are affected by the presence of the boat, but it explicitly excludes any backaction so that the ocean waves have \emph{no effect at all} on the boat. This we find to be an absurdity! Surely any reasonable model of the environment must permit a great disturbance to the system so that the initial state is scrambled. 

The difficulties faced by \eqref{3} do not stop there. It is clear that in a normal experimental setup in which observable $O$ is measured, a device of a very particular nature must be constructed. If one wishes to measure $O$ instead of $O'$, then two different devices must be constructed. For example, it is evident from experiment that devices which make position measurements are dramatically different from those devices which perform momentum measurements. This obvious fact is missing from \eqref{3}. Indeed, the correlation introduced by the apparatus alone works for all observables, as we have noted, and it is, according to \eqref{3}, the environment which finally selects the ``observed" observable (for lack of a more concise way of stating it). It is suspected that the environment exhibits statistical randomness and furthermore it is assumed that the state of environment is outside of the experimenter's control. Thusly we should expect, in a best case scenario, that the experimenter does not decide which observable is made manifest in a particular measurement. The notion of ``constructing a position measuring apparatus to measure positions" is well documented but does not have a place in the decoherence program. Even worse is the possibility that different observables may be selected in different trials of an experiment because of changes in the environment.

In the decoherence program, one tacitly assumes that $\ket{o_i}$ in \eqref{3} are the eigenstates of an observable. However, without an adequate model of the unitary evolution in \eqref{3} or \eqref{3''} one cannot claim to know the states $\ket{s_i}$. \emph{There is currently no mechanism within the decoherence program that guarantees that states $\ket{o_i}$ will be the eigenstates of an observable}. Assume at this point that it can be demonstrated that $\ket{o_i}$ are the eigenstates of an observable. For something definite and without loss of generality, one might imagine that each $\ket{o_i}$ corresponds to a discretized position eigenstate, $\ket{x_i}$ referring to ``box" $i$. Now, it might seem that the formalism indicates that interaction with the environment has lead to the localization of system $S$ to within a particular box $\ket{x_i}$. This conclusion is false because a superposition of many eigenstates remains. We conclude that, despite the environmental interaction in \eqref{3}, the resulting state still suffers from the superposition problem. Indeed, no superpositions have been destroyed nor have any states with definite classical properties arisen. Outside of a relative state or ``many-worlds" approach (we will discuss this possibility in section \ref{MW}), the state \eqref{3} does not lend itself to interpretation.

As we have previously mentioned, if the environment is a quantum system (all interactions are unitary), then the final state remains pure, and the interpretation of such a state requires the relative state formalism. Adherents to the decoherence program, who apparently reject this possibility, argue for tracing over the environmental degrees of freedom \cite{ZurekII}. If the interaction with the environment is responsible for generating state \eqref{3}, then what is responsible for transforming \eqref{3} into the reduced state, \begin{equation}\rho_S=Tr[U_{SE}\ket{S_{in}}\otimes\ket{\epsilon_{in}}\bra{S_{in}}\otimes\bra{\epsilon_{in}}U^T_{SE}]\label{4}?\end{equation} The answer most often cited is that of \emph{environmental ignorance}. Namely, the environment is not measured or the experimenter lacks control over the environment. The mention of ``measurement" or ``ignorance" in an argument for the trace procedure is unacceptable. Doing so reintroduces measurement as a fundamental process, which is the main difficulty with the Copenhagen interpretation that a new interpretation must address. How can the mere ignorance of the measurer (the experimenter) manifest itself as a physical reduction of the system from pure state \eqref{3} to mixed state \eqref{4}? This statement indicates that the mental state of the experimenter has physical bearing experiments. Surely this is spookier than the state reduction at the hands of a measuring device in the Copenhagen interpretation. At least in the latter case, a physical interaction (between measuring device and system) may be blamed! 

Tracing is explicitly nonunitary. Recall that, as we have argued in section \ref{CI}, nonunitary behavior is (at least in principle) distinguishable from purely unitary behavior. Let us extend the argument of section \ref {CI} to the general case of arbitrary nonunitarity, including decoherence from environmental interaction. Consider a total system, which consists of a system of interest $S$, an observer $A$, and the environment $E$. We will assume that the observer $A$ is able to ``measure" system $S$, but is incapable of measuring the environment, as per the argument for tracing in the decoherence approach). Equivalently, we may assume that $A$ chooses to ignore the environment (also per the argument for tracing in the decoherence approach) Regardless of how one wishes to argue for properties of $A$, the final result is that $A$ is assumed capable, via interaction, of reducing pure density matrices to mixed density matrices. Since the environment is considered part of the system, the total system may be regarded as closed (there is nothing more external to the total system with which its component systems, $S$, $A$, or $E$ interact). If all closed systems evolve unitarily, then the state of the total system must be \begin{equation}U_{SAE}\ket{S_{in}}\otimes\ket{A_{in}}\otimes\ket{\epsilon_{in}}=\sum_ik_i\ket{o_i}\ket{a_i}\ket{\epsilon_i}\label{5}\end{equation} which is pure. If there is a reduction from pure to mixed state induced by any of the interactions, either with the observer $A$ or possibly with the environment, then the final state will not be pure, but will instead be described by a mixed density matrix $\rho$. It should be clear that the details of how the nonunitarity arises are not important for this argument. It will go through as long as some type of nonunitarity gets introduced in the final state. To complete the argument recall from Section I, that pure or mixed final states form mutually exclusive alternatives as they give rise to different experimental results. Thus, we find that the assumption of unitary evolution of closed systems is \emph{incompatible} with any nonunitary evolution whatever. One might be tempted to evade this conclusion by arguing that the reduction from \eqref{3} to \eqref{4} is not ``real." That is to say, the system and environment truly are in state \eqref{3}, and \eqref{4} merely arises as an effective description. However, it is evident that \eqref{4} cannot replicate \eqref{3} for all observables, as there is a loss of phase relationships in the transformation from \eqref{3} to \eqref{4}, and correspondingly a loss in possible interference effects. One might hope that for a limited class of observables \eqref{4} does replicate \eqref{3}. Under which circumstances this may be the case is certainly of pragmatic interest. It cannot, of course, be of fundamental theoretical interest, as the fundamental object is the state \eqref{3}, nor can we be concerned with a limited class of observables at a fundamental level. Note that without a physical state reduction, the evolution is purely unitary, and thus is an example of a relative state/MW theory, the discussion of which we withhold until the section \ref{MW}.  

\subsection{Initial Entropy Problem}
\label{IEP}
It is evident that a realistic environment has a Hilbert space with a large number of dimensions; however, insight can be had by considering much simplified models. In particular, we have introduced an environment whose approximate base may be taken to be $\Gamma=\{\epsilon_{in},\epsilon_{j}\}$. It is evident that $\Gamma$ functions as an approximate base because each state in $\Gamma$ corresponds to a macroscopically distinguishable situation, $\ket{\epsilon_{in}}\rightleftharpoons$ ``the environment has not interacted with $S$," $\ket{\epsilon_j}\rightleftharpoons$ ``the environment recognizes value $o_j$ for observable $O$," and must therefore satisfy $\braket{\epsilon_i}{\epsilon_j}\propto\delta_{ij}$; $\braket{\epsilon_j}{\epsilon_{in}}\approx 0$. We now observe that for the decoherence program to replicate the Born rule, the initial state of the environment must be $\epsilon_{in}$, see equation \eqref{3}. Of particular note is that the initial environment state \emph{cannot} involve a superposition of $\ket{\epsilon_{in}}$ and $\ket{\epsilon_j}$, because such an initial superposition immediately spoils the interpretation of \eqref{3}, and thereby produces a mismatch between the predictions of decoherence and the Born rule. We conclude that the decoherence interpretation requires a superselection rule on the initial environment state which filters superpositions in order to guarantee agreement with experiment. Equivalently, it may be said that decoherence requires fine-tuning because only one state out of many is allowable as an initial state. One may na\"ively believe that one must model with realistic environments. However, in any realistic environment, there are a very large number of basis states. As a result, the fine-tuning is exacerbated because one is selecting $\epsilon_{in}$ (more precisely the equivalent of $\epsilon_{in}$ in a realistic environment model) as the only viable initial environmental state out of many. 

The requirement of such a superselection rule is both unnatural and unprecedented. The initial states of classical systems are free to explore the whole of phase space. It is expected that quantum systems should exhibit the same freedom. The restriction of the initial state to non-superposed states (i.e. the initial state must be \emph{ultra pure})\footnote{States that are described by a Hilbert space element are referred to as \emph{pure}. We introduce terminology such that states that are single basis elements in a preferred basis are \emph{ultra pure}.} indicates that the initial entanglement entropy is low, since there are many more states that correspond to entangled states which must be excluded as possible initial apparatus states. Accordingly, the initial state is a highly ordered state. To phrase this colloquially, consider a formalism describing the air in the room. While it is possible that the initial state of the system corresponds to a macrostate in which all the air was found in one corner of the room it is extremely unlikely. If the formalism \emph{demanded}, on the basis of consistency, that the initial system state be such an unlikely configuration we would hardly give it credence. Without an additional mechanism that selects the initial states, this interpretational requirement enters the theory as an ad hoc, and unjustifiable, assumption. We refer to this deficiency as the \emph{Initial Entropy Problem}, and note here that it appears in other formulations of quantum mechanics. 

A similar question with regard to an initially low entropy and the need for fine-tuning arises in cosmological considerations. Namely, the second law of thermodynamics suggests that the initial entropy state of the universe is extremely low. If this is so, then fine-tuning is apparently needed to account for this low entropy initial state. See \cite{Carroll1} for further details.

\subsection{Observer Energy Problem}
\label{OEP}
An essential point is that the system of interest, the measuring apparatus, and the environment must necessarily form a \emph{closed system}. Well known is that this demands that the evolution of the full system is unitary. Often overlooked, however, especially within the context of measurement problem resolutions, is the fact that closed systems must also conserve energy. Does the decoherence formalism satisfy this well established dictum? Let us demonstrate that it does not. Recall that energy conservation in quantum mechanics demands that the Hamiltonian is time independent. Notice also that the evolution of quantum systems thus far observed in experiment has two regimes of evolution. The first regime is the ``regular" unitary regime wherein the system of interest $S$, the apparatus $A$, and the environment $E$ evolve independently. In the second regime, herein referred to as the ``decoherence regime," $S, A$, and $E$ become strongly coupled. Indeed, the coupling is assumed so strong that the time evolutions of the individual systems become negligible in comparison. Call time $t^*$ the \emph{decoherence-on time}, and let it be the time that decohering effects become important. Then the evolution of the full system before $t^*$ is governed by $H_{S}, H_{A}, H_E$. The system of interest, the apparatus, and the environment evolve independently in the regular unitary regime. After $t^*$ the systems interact via $H_{SAE}$ and the systems enter the decoherence regime. It is straightforward to see that the full Hamiltonian is \begin{equation}\left(H_S+H_A+H_E\right)\epsilon(t^*-t)+H_{SAE}\epsilon (t-t^*),\label{timed}\end{equation} where $\epsilon (x-y)$ is the Heaviside step function. It is clear that equation \eqref{timed} has explicit time dependence, and therefore does not conserve energy. It is also the Hamiltonian of a closed system, and thus violates the conservation of energy for a closed system. An objection may be raised because of the sharp turn on time $t^*$, but it is easy to see that even if the turn on were smooth, in so much as two regimes of evolution are present, some explicit time dependence is necessary, and thus the argument would go through. As a matter for future work, it may be possible to begin with a single time independent Hamiltonian whose evolution on the full system exhibits the two observed regimes. We caution, however, that this approach for resolution is difficult because one would need to interpret from the dynamics of the system the decoherence-on time $t^*$, which appears to be a degree of freedom in a normal experimental setup (i.e. an experimenter can allow decoherence to occur at any time she wishes). Thus, we see that this approach has features of superdeterminism. Since the apparatus and environment function effectively as an observer and in accord with this, we refer to this deficiency as the \emph{Observer Energy Problem}.

\subsection{The Tails Problem}
The action of tracing out the environmental degrees of freedom from the full density matrix produces an approximately off diagonal density matrix. This density matrix, the ``system density matrix," may therefore be interpreted as describing an approximate mixed state. However, the off diagonal terms do not vanish completely for any finite time. As a consequence, the system density matrix is only mixed to the extent that the off diagonal terms may be neglected. In a position base (if applicable) the nonzero off diagonal terms extend to spatial infinity and as a result constitute ``tails" of the distribution \cite{Lewis}. Although no quantitative difficulty arises from the presence of tails, they complicate the interpretation by demanding that one make a judgment of size when the formalism does not supply a scale. That is, when are the off diagonal terms sufficiently small in comparison to the diagonal terms so that the density matrix is mixed? We mention here in passing that the \emph{tails problem} is not restricted to the decoherence formalism, but appears also in objective collapse theories \cite{Schlosshauer1} \cite{Schlosshauer2}.

\section{Relative State/Many Worlds Interpretation}     
\label{MW}
The relative state/many worlds (RS/MW) interpretation represents a radical departure from either the Copenhagen interpretation or decoherence \cite{HughEverett1957}. In this approach, it is assumed that \emph{all} evolution is unitary. No collapse, projection, nor `trace-type' behavior ever occurs. Instead, Von Neumann measurement of the first kind, which permits the formation of a correlation between a system and an apparatus via unitary evolution alone is taken to be sufficient for describing experiment. Of course, the state of perfect correlation between system and apparatus \emph{does not} select a particular measured value. The state is one of superposition among perfectly correlated terms (i.e. it is of the form $\sum_i \ket{i}_{system}\ket{i}_{apparatus}$). A point of contention arises when one is forced to interpret the standing superposition. It seems that the most natural interpretation is that all allowed correlations persist, thereby generating a set of ``quantum worlds." While this might be at odds with some innate sense of ``simplicity," the resulting theory has fewer assumptions than other approaches, and so is quantitatively simpler. The RS/MW interpretation seems at first quite promising because of the great elegance in eliminating assumptions about intermittent non-unitary evolution; however, there are assumptions necessary for the interpretation which, upon closer examination, are severely \emph{ad hoc} \cite{Tegmark}. Let us briefly review the relative RS/MW interpretation. We once again begin by considering a system $S$ in state $\ket{S_{in}}$. The evolution in all circumstances is purely unitary and so we need only focus on measurement for a complete understanding of the interpretation. We introduce an apparatus $A$, with initial state $\ket{A_{in}}$. The interaction between the system and the apparatus is as follows \begin{equation}U_{SA}\ket{S_{in}}\otimes\ket{A_{in}}=\sum_jk_j\ket{o_j}\otimes\ket{a_j},\label{VNM1}\end{equation} 
where $\ket{S_{in}}=\sum_jk_j\ket{o_j}$ and where $\ket{o_j}$ corresponds to ``the system has value $o_j$ for observable $O$," and $\ket{a_j}$  corresponds to ``the apparatus determined that the system has value $o_j$ for observable $O$." It is evident how \eqref{VNM1} yields the interpretation that the apparatus has measured a particular value for the observable since all the terms demonstrate perfect correlation between the system state and the apparatus state. Notice also that the superposition survives and so \eqref{VNM1} must be interpreted as a set of quantum worlds \cite{HughEverett1957} \cite{Bell}.

It appears, at first glance, that \eqref{VNM1} corresponds to a measurement of observable $O$. However, it is evident that it suffers from the \emph{preferred basis problem} (see section \ref{Decoherence}). That is, under a change of base from the eigenstates of observable $O$ to the eigenstates of a different observable $O'$ \eqref{VNM1} maintains perfect correlation. Thus, one \emph{cannot} claim that \eqref{VNM1} corresponds to a measurement of any particular observable. This difficulty, of course, is well known and in fact spurred on the decoherence approach, which we have argued in section \ref{Decoherence} fails to adequately address these issues. 

The RS/MW interpretation faces additional deficiencies. In the terminology of the previous section, it is said that the RS/MW suffers from both the initial entropy problem and the observer energy problem. Let us review these problems in the context of the RS/MW interpretation. The interpretation requires that a Von Neumann measurement (of the first kind), such as \eqref{VNM1}, be realizable. In turn, such interactions require particular initial states for the measuring apparati. That is, the measuring apparatus \emph{must} be in state $\ket{A_{in}}$. If the initial state of the measuring apparatus is any other state or a linear superposition of $\ket{A_{in}}$ and other states, then the interpretation fails. Since the initial state must be very precisely fixed, it is in a very low entropy state, yet, no mechanism for establishing such low initial entropy is provided (see section \ref{Decoherence}). 

The RS/MW interpretation suffers from the observer energy problem (see \ref{OEP}). The full system, composed of apparatus (observer) and system of interest, is a closed system. There is, initially, assumed to be no interaction between the system of interest and the apparatus; however, at a finite time $t^*$ the interaction must be ``turned on." This induces a time dependence in the (full) Hamiltonian, which implies nonconservation of energy. This is inconsistent with the well verified principle of energy conservation together with the assumption of a closed system.  

Finally, defining probabilities in the RS/MW interpretation faces challenges. It is, strictly speaking, impossible to derive the Born rule and the corresponding probability interpretation from the RS/MW formalism because one assumes only unitary evolution on an appropriate Hilbert space, which contains no inherent notion of sample space, frequency, degree of belief, nor any other quantity that may be bridged to probability theory without further assumption. Claims that the Born rule have been ``derived" from a purely Everettian approach (without additional assumptions) are faulty \cite{Carroll2} \cite{Carroll3}. The possibility of arguing for the Born rule as a unique norm on quantum worlds remains open \cite{DD} \cite{Wallace1} \cite{Wallace2} \cite{Wallace3} \cite{Wallace4} \cite{Saunders1} \cite{Saunders2}. If superpositions always remain, then the number of quantum worlds forms a continuum. To define probabilities on the continuum requires an appropriate probability measure. Whether this can be defined, is unique, and if it follows from the dicta of the approach, are open issues. It is not even clear how the quantum worlds should be interpreted. For example, should the frequentist inference be used or must some other inference scheme be used/developed \cite{Kent}?

 \section{Bohmian Mechanics}

Bohmian mechanics attempts to resolve not only technical difficulties concerning measurement, but also the conceptual difficulties facing quantum mechanics by postulating that particles follow a trajectory (have a position at all times). The trajectory is dictated by a guiding equation, which is written in terms of a wavefunction \cite{Bohm1} \cite{Bohm2} \cite{Bohm3}. The wavefunction satisfies the standard Schr\"odinger equation. This approach is not an interpretation, but instead a competing theory \cite{Bell}. 

Let us proceed by reviewing the treatment of a single point-like system. The wavefunction satisfies the standard Schr\"odinger equation, which we do not state. The guiding equation is 
\begin{equation}m\frac{d\mathbf{x}}{dt}=\hbar\phantom{i}\mathrm{Im}\left(\frac{\nabla\psi (\mathbf{x}(t),t)}{\psi(\mathbf{x}(t),t)}\right)\label{BM0},\end{equation}
where $m$ is the mass of the particle. The guiding equation is first order in time. The resulting theory is fully deterministic. Notably, probability does not appear as a fundamental constituent of the theory. Let us raise our first objection. As defined by the Schr\"odinger equation and the guiding equation, \eqref{BM0}, this theory is \emph{not} equivalent to standard quantum mechanics. Its predictions may differ wildly from those of the Copenhagen interpretation, for example, and is as a consequence, strongly at odds with experiment. In order to guarantee agreement with experiment one must impose the \emph{quantum equilibrium hypothesis}. It states that the probability distribution for positions must be given by $|\psi(\mathbf{x},t_0)|^2$. Once this distribution is fixed at one time $t_0$ it will be preserved under time translation \cite{Bohm1} \cite{Bohm2} \cite{Bohm3}. It is uncertain (and not specified by the theory) whether the quantum equilibrium hypothesis is always fulfilled or if nonequilibrium states exist in nature (if only for short times) \cite{Bell}.

The guiding equation indicates that for real-valued spatial wavefunctions the velocity of the Bohmian particle vanishes. This predicts physical phenomena. To demonstrate that this feature makes predictions in conflict with experiment we need only consider the ground state of the hydrogen atom. The spatial wavefunction corresponding to this state is real-valued and the corresponding Bohmian particle velocity vanishes. Therefore, the ground state of the hydrogen atom would form a permanent electric dipole. In addition to greatly altering the interactions in (single atom) hydrogen gas, the permanent dipoles could be aligned in weak electric fields, producing photon flux as they de-excite from the higher energy anti-aligned state. None of these phenomena are observed, and thus it must be concluded that Bohmian mechanics is inconsistent with experiment \cite{Dick2012} \cite{Jung2013}. 

Like several other interpretations we have considered, measurement in Bohmian mechanics relies on the use of a Von Neumann measurement of the first kind. That is, it assumes that the system of interest $S$ and the apparatus $A$ originally have independent wavefunctions $S_{in}(\mathbf{x})$ ($\mathbf{x}$ is the position of a particle) and $A_{in}(y)$ ($y$ is the position of the pointer), respectively, and that during the measurement process the evolution is as follows \begin{equation}U_{SA}[S_{in}(\mathbf{x})A_{in}(y)]=\sum_jk_j\psi^{(S)}_j(\mathbf{x})\phi^{(A)}_j(y)\equiv \Psi(\mathbf{x},y,t)\label{VNM2}\end{equation} 
where $S_{in}(\mathbf{x})=\sum_jk_j\psi^{(S)}_j(\mathbf{x})$, and where $\psi^{(S)}_j(\mathbf{x})$ is the eigenstate of observable $O$ with eigenvalue $o_j$ and $\phi^{(A)}_j(y)$ is the wavefunction of the ``pointer position" corresponding to eigenvalue $o_j$. In the relative state/many worlds interpretation equation \eqref{VNM2} would be said to describe a measurement of observable $O$, where the quantum worlds remain in superposition. Within Bohmian mechanics, the wavefunction generates the equations of motion for the particles, 
\begin{eqnarray}m_x\frac{d\mathbf{x}}{dt}&=&\hbar\phantom{i}\mathrm{Im}\left(\frac{\nabla_x \Psi (\mathbf{x},y,t)}{\Psi(\mathbf{x},y,t)}\right)\label{BM1}\\m_y\frac{dy}{dt}&=&\hbar\phantom{i}\mathrm{Im}\left(\frac{\nabla_y \Psi (\mathbf{x},y,t)}{\Psi(\mathbf{x},y,t)}\right)\label{BM2},\end{eqnarray} where $m_x$ is the mass of the particle system and $m_y$ is the effective mass of the pointer. It is then supposed that $\phi^{(A)}_j(y)$ are sharply peaked and at late times are sufficiently widely separated that an observation of $y$ allows one to deduce which state (packet) $\phi^{(A)}_j(y)$ is occupied, which from equation \eqref{VNM2} allows one to learn which system state (packet), $\psi^{(S)}_j(\mathbf{x})$, is occupied, so that one has effectively ``measured" observable $O$ and that the effective wavefunction for the system is just $\psi^{(S)}_j(\mathbf{x})$. Finally, it is argued that the empty states will not again interfere because the system and apparatus will have already coupled to many other degrees of freedom. See \cite{Bohm2} for a more detailed presentation.

 There are several important objections to this description of quantum measurement. Firstly, equation \eqref{VNM2} suffers from the initial entropy problem --- introduced in \ref{IEP}. The initial apparatus state cannot be in a superposition or the interpretation of equation \eqref{VNM2} fails as a ``measurement." The initial state is, consequently, a very low entropy state, and without a mechanism to reduce the (entanglement) entropy, such as intermittent nonunitary behavior it is exceedingly unnatural that the apparatus be initially in such a state. Secondly, equation \eqref{VNM2} suffers from the observer energy problem --- introduced in \ref{OEP}. Namely, it is essential that a Von Neumann measurement Hamiltonian be induced between the system of interest and the apparatus at some finite time $t_0$. Therefore, the Hamiltonian is endowed with explicit time dependence and accordingly energy is not conserved. As a result, either the system of interest + the apparatus cannot be regarded as a closed system or we must accept a violation of energy conservation. Finally, it is tacitly assumed that the pointer position $y$ may simply be ``observed" or, if you will, read from the dial. We say that, in the Bohmian mechanics approach, the pointer position is ``observable without disturbance," or simply, OWD. It is evident that in the treatment there is \emph{complete symmetry} between $\mathbf{x}$ and $y$. That is, both the system and the apparatus are given a full quantum description. If the pointer position may be observed without disturbance, then why is the system (particle) not also OWD? It is claimed that the pointer position is a near classical variable, but it is not described how this near classical behavior arises nor is it described how the pointer position becomes observable (it is still described by a wavefunction). More precisely, how does one \emph{quantitatively} adjust the treatment of the subsystem $y$ so that $y$ is near classical? A na\"ive response may be to take the pointer very massive so that wavefunction spreading is negligible. However, there is no justification for taking the initial wavefunction to have a single peak, and even with the quantum hypothesis satisfied many initial pointer positions will not coincide with the peak. Additionally, a massive pointer fails to introduce a distinction (classical versus quantum) between the system and the pointer, because the system too has nonzero mass. What we mean by this is that the mass is continuously variable, and properties that change as functions of the mass are also continuously variable, as a consequence if the pointer is \emph{approximately} OWD, then the system is also OWD, albeit to a worse approximation. We argue more generally as follows. From a fundamental standpoint the system and apparatus are (and indeed \emph{must} be) treated symmetrically --- both are fully quantum systems. If the pointer may be observed ``by simply looking at it" (is OWD), then by logical consistency the system must also be OWD. The claim that quantum systems follow trajectories that are OWD is currently outside experimental justification. 
 
Bohmian mechanics, as defined by the Schr\"odinger equation and the guiding equation \eqref{BM0} does not delineate a regime for a quantum to classical transition. Without modification (or more likely additional dynamical rules), nothing prevents Bohmian mechanics from predicted large scale quantum effects, which are obviously not observed. The most common remedy is to introduce decoherence into the theory, which allows one to compute necessary quantities such as decoherence times. We argue against decoherence as an independent interpretation in Section \ref{Decoherence}. A Bohmian-decoherence hybrid theory retains the deficiencies of both, at least to the extent that Von Neumann measurement of the first kind remains an essential constituent.

Bohmian mechanics, which incorporates interactions among the full configuration space (particles interact directly even when spatially separated) cannot readily be made relativistic. It is unclear how, or importantly ``if," one can realize a relativistic version of the Bohmian theory that is not also immediately at odds with experiment. Related to this problem is the question of producing a ``Bohmian field theory" that accounts for particle creation and annihilation. These both constitute matters of future work.

\section{Modal Interpretation}

The modal interpretation, or more accurately, ``modal interpretations" disposes of the standard eigenstate-eigenvalue link (a system has a definite property $o_i$ \emph{iff} it is eigenstate $\ket{o_i}$) and instead propose that a system always has a set of definite properties, referred to as the \emph{value state}, which is independent of the \emph{dynamic state}, which always evolves according to the Schr\"odinger equation \cite{Dieks} \cite{vFraa} \cite{Bacciagaluppi}. It is obviously necessary to introduce a mechanism that selects which properties have definite values and at what times. This rule for ascription of definite properties is called an \emph{actualization rule}. Currently this ``interpretation" is incomplete because no definitive actualization rule has been agreed upon. Instead, \emph{many} possible actualization rules exist and many have been proposed. For example, one may take the possible value states to be elements in an orthonormal basis of the Hilbert space of the system and a probability density is defined on these values in accordance with the Born rule. How is the orthonormal basis selected? It is \emph{not} selected from the unitary formalism. Consequently, we see that the modal interpretation suffers from a preferred basis problem. One may also permanently select a preferred basis, for example, a position base. The result is a theory \emph{resembling} Bohmian mechanics; however, the formalism does not generate an evolution equation for the position state (the guiding equation, \eqref{BM0}, or some equivalent must be supplied). One might take energy or momentum as a preferred base, but such possibilities have not been explored. This approach is currently incomplete. Indeed, depending on how one proposes to complete it the approach may have an enormous possible number of ontologies and subsum many different interpretations. It is for this reason that we reluctantly deem the modal approach an ``interpretation."

Na\"ive attempts to complete the modal interpretation give rise to familiar problems, for example, the preferred basis problem. Similarly, modal interpretations often makes use of the decoherence mechanism, which we argue in Section \ref{Decoherence} gives rise to \emph{initial entropy} and \emph{observer energy} problems. The resulting modal interpretation would also suffer the aforementioned deficiencies. Finally, eliminating the eigenstate-eigenvalue link raises the possibility of measurements of property $O$ that reveal value $o_i$, but output states different than $\ket{o_i}$. More precisely, without an eigenstate-eigenvalue link measurements of $O$ that reveal value $o_i$ may output states of the form $\sum_ik_i\ket{o_i}$, which indicates that a second measurement may reveal a value for $O$ different than $o_i$. The occurrence of such phenomena is easily experimentally distinguished from quantum mechanics under the Copenhagen interpretation, and has hitherto been unobserved.     

\section{Objective Collapse Theories}
\label{OC}
Objective collapse theories purport that collapse is a physical phenomenon. Often the collapse is assumed spontaneous and may be induced by a nonlinear term in the Schr\"odinger equation \cite{Ghose} \cite{GRW1} \cite{GRW2}. These theories do not constitute a reformulation or interpretation of quantum mechanics, but instead make dramatically different experimental predictions \cite{GRW1}\cite{GRW2}. It is precisely for this reason that these theories are non viable. While more advanced searches for specific types of nonlinearities are certainly warranted, it is easily demonstrated that objective collapse theories with spontaneous collapse, such as, for example, the Ghirardi-Rimini-Weber theory, are contradictory to known experimental results. Consider the double slit experiment. The initial state of each particle must be in a (near) momentum eigenstate of the same momentum as an initial state or the interference pattern will be washed out. If a spontaneous collapse occurred with a given frequency, then surely some of the particles would suffer a collapse during the travel between the source and the detection screen. Indeed, some fraction of them will collapse at or immediately before the double slits. The interference pattern, after many trials, would be disturbed. Notice that the particles need not collapse to a position eigenstate, any nonlinear evolution or stochastic jump, occurring even with low probability, would manifest itself by deforming the interference pattern. No such deformation has been observed. 

It is possible that spontaneous collapses occur with an extremely low probability. However, the low probability of their occurrence dictates that they cannot play an essential role in resolving the measurement problem. Again, consider the double slit experiment. Very few spontaneous collapses occur en route, yet, an overwhelmingly large number of particles produced by the source are detected by the detection screen in near position eigenstates. An additional mechanism must be responsible for the abrupt collapse that occurs at the detection, which is not described at the present time by any objective collapse theories. 

\section{Consistent (Decoherent) Histories}

The consistent (decoherent) histories approach to the measurement problem proposes that the fundamental ontology of closed systems is a ``history." Specifically, at each time $t_i$ in a series of times $t_i=\{t_0,t_1,...t_n\}$ one associates an exhaustive set of projection operators $\{P_j^{(t_i)}\}$ ($\sum_jP_j^{(t_i)}=I$). Observe that for each time $t_i$ the set $P_j^{(t_i)}$ forms a partition of the identity in the index $j$. We highlight this property because it will prove vital for interpretation. One constructs a history, $Y$, as a time ordered sequence of projection operators \begin{equation}Y=P_{j_0}^{(t_0)}P_{j_1}^{(t_1)}P_{j_2}^{(t_2)}\cdots P_{j_n}^{(t_n)},\end{equation} and ascribes to it the interpretation that the system has property $P_{j_0}$ at time $t_0$, property $P_{j_1}$ at time $t_1$,... and property $P_{j_n}$ at time $t_n$. Since the projection operators form a partition of the identity, the associated properties are mutually exclusive. It is evident that this interpretation implies that a system has a well defined property only at the times $t_i$. At intermediary times no information is known about the system. One defines a \emph{chain operator} as follows, \begin{equation}P_{j_n}^{(t_n)}U(t_n,t_{n-1})\cdots P_{j_2}^{(t_2)}U(t_2,t_1)P_{j_1}^{(t_1)}U(t_1,t_0)P_{j_{0}}^{(t_0)}.\end{equation} So that the system evolves unitarily in between ``projections." We now wish to assign a probability to a history. The key demand that we require is that the probabilities should be additive, so that the probability of the system following history $Y$ ($\equiv P(Y)$) \emph{or} the probability of the system following history $X$ ($\equiv P(X)$) is $P(Y)+P(X)$. It is evident that in general the additivity property fails because of the unitary evolution. On what set of histories will this property hold? It is straightforward to demonstrate that a sufficient condition for additivity between two histories $Y$ and $X$ is \begin{equation}Tr\left[C_Y \rho_{in} C^\dagger_X \right]=0,\label{CH0}\end{equation} where $C_Y$ and $C_X$ are the corresponding chain operators of histories $Y$ and $X$ and $\rho_{in}$ is the initial state of the system. If the condition \ref{CH0} holds for any two histories in a set, the set is referred to as a \emph{realm}. One then defines the probability for a particular history $Y$ to be followed within the realm to be \begin{equation} P(Y)\equiv Tr\left[C_Y \rho_{in} C^\dagger_Y \right]. \end{equation} We note in passing, as it will be useful for our purposes, that the set of single-time histories (a single projector is inserted at time $t_0$) form a realm, as is readily proved directly from the consistency condition \eqref{CH0} \cite{Griffiths2002} \cite{G1} \cite{G2} \cite{O1} \cite{G-MH1} \cite{G-MH2012}.

\subsection{Realm Selection Problem}
A deficiency arises in this formulation because of the dependence on choosing a realm. Consider, as an example, a system composed of a single particle. It is straightforward to demonstrate that the set of histories such that the particle has a well defined position $\mathbf{x_0}$ at $t_0$ forms a realm. In a like manner, one may show that the set of histories such that the particle has a well defined momentum $\mathbf{p_0}$ at $t_0$ also forms a realm. These realms are obviously mutually inconsistent because the position and momentum operators fail to commute. This system thus has (at least two) realms. We now ask a question of the formalism ``what is the probability that the system has a position (as opposed to a momentum)?" It is clear that the formalism alone provides no answer to this question. Immediately we see that the consistent histories formalism demands a realm-dependent reality -- only after a realm is selected can one compute relative probabilities. How is a realm selected? It is apparent that realm selection is a \emph{physical} process (i.e. in practice a realm must be selected by a choice of experiment or the design of the measuring apparatus), yet a description of this process is excluded by the the consistent histories formalism. This is in very strong analogy with the Copenhagen interpretation, which introduces a physical process, measurement, yet a description of this process is excluded from the theory. Indeed, the consistent histories formalism contains a mere repackaging of the measurement problem as it arises within the Copenhagen interpretation. 

Call a theory \emph{incomplete} if there exists a statement which is undecidable from within the theory (the theory makes no prediction), but which is decidable through experiment. Consider a system that has at least two realms (we will prove later that all systems have at least two realms). We will refer to them as realms $A$ and $B$. Consider the statement ``the system of interest follows a history within realm $A$ with probability $\mathcal{P}(A)$." An equivalent statement holds for realm $B$. We will refer to this as \emph{statement $\alpha$}. It is clear that this statement is undecidable through the consistent histories formalism because it asks about a relative probability between different realms and the formalism only assigns probabilities between histories in a fixed realm. Statement $\alpha$ is decidable through experiment. If experiments are capable of detailing the history that a system follows, then one may compute $\mathcal{P}(A)$ from the ratio $n_A/N$, where $n_A$ is the number of observed histories that lie in realm $A$ and $N$ is the total number of observed histories. Therefore, statement $\alpha$ is decidable through experiment, but undecidable through the formalism, and thus the formalism is incomplete. 

An adherent to consistent histories may object to the aforementioned conclusion on the basis that the realm is \emph{chosen} by the experimenter, so that the ``probability for a given realm" is meaningless. Let us demonstrate that the probability for a given realm is necessary for consistency and argue that it is an obvious quantity to be deduced from experiment. Consider as a system, a system of interest, an apparatus, and an experimenter. A key point is that this system is closed. There are two notable consequences of closure. Firstly, the consistent histories formalism deems that it must follow a particular history (in some realm). Secondly, there is \emph{nothing} outside the system to enact realm selection. It is straightforward to see that there are only two ways to guarantee the consistency of these two statements. Since a realm must be selected, either the system states/dynamics must be such that there is a single realm (so that there is no choice of realm) or a relative probability among realms must be included in the theory (so that realms are random but occur with different weights). We will first show that there are at least two realms for this system. Consequently, we will be forced to conclude that a relative probability among realms must be included in the theory. As a first step in proving that this system has at least two realms notice that if the system of interest is a quantum system, then it has at least two noncommuting hermitian (observable) operators, without loss of generality we will take them to be $\hat{\mathbf{x}}$ and $\hat{\mathbf{p}}$. Then two realms may be formed by considering the single time realm with well defined position and the single time real with well defined momentum. Let us now consider the full system with apparatus and experimenter included. Apart from the placement and choice of projection operators, the evolution type in the chain operators is unitary, and unitary evolution  cannot select a base, as per the preferred basis problem, it is clear that there exists a single time realm where the projector projects onto a definite value of position ($\mathbf{x_0}$ for definiteness) and the states ``the apparatus has found $\mathbf{x_0}$," and ``the experimenter has observed $\mathbf{x}_0$." Similarly, there exists a second single-time realm where the projector projects onto a definite momentum $\mathbf{p}_0$. Since the operators fail to commute, these are two distinct realms for a generic system, as we wished to demonstrate. The claim of incompleteness continues to hold. 

\subsection{Inconsistency with nonexistence of null result measurements}
There are two features, apparent from experiment, that will prove valuable in highlighting a further deficiency in consistent histories. Firstly, the times at which measurements occur appear to be determined by the experimenter herself. We do not use the term \emph{measurement} here to refer to Copenhagen style measurement, but is instead just a name for a the physical process that occurs during an experiment. That is to say that there appears to be a complete freedom of choice in deciding when the measurement will occur. It is difficult to proceed if this freedom is not exhibited, as the resulting theory is superdeterministic. In the following discussion we ignore the possibility of a superdeterminism. Secondly, a measurement of property $O$ \emph{always} reveals that the system is in an eigenstate of $\hat{O}$. We refer to measurements of a property whose result is that the system does not have the aforementioned property as null result measurements (not to be confused with null measurements). With this terminology, one may say that null result measurements do not exist. We may now state a main result. \begin{quote}The statement ``The consistent history followed by a system is $Y$" is false or null result measurements exist.\end{quote}
\emph{Proof}: Assume that the consistent history followed by a system $S$ has is determined to be $Y$. An experimenter makes a measurement at any time $t_A\neq t_i$, which is not one of the sequence times. If a result is obtained, then the true consistent history was not $Y$, since $Y$ indicates that the system has a well defined property at the sequence times, and this system has a well defined property at the additional time $t_A$. If a non result is obtained, then null result measurements must exist. Being able to determine the history of a system is essential for comparing this formalism to experiment. Indeed, if the history is not known nor is the probability. This result indicates that the consistent histories approach is either in stark disagreement with experiment or is a non-predictive theory. 

Additional arguments against the consistency of the consistent histories approach may be found in \cite{Dowker1}, \cite{Dowker2}, \cite{Sudarsky2014}, among others.

\section{Transactional Interpretation}
\label{TI}
The transactional interpretation strongly derives from the Wheeler-Feynman absorber theory, which we review briefly \cite{Cramer1}. It is well known that Maxwell's equations admit both \emph{retarded} and \emph{advanced} solutions, in which effects are delayed or advanced by the light travel time. Classically, the advanced solutions are disregarded as nonphysical. In quantum field theory, one cannot simply banish the advanced solutions by fiat. Indeed, the correct choice of Green's function (propagator) is the Feynman propagator, which is fully time symmetric because it is composed of an equally weighted sum of retarded and advanced propagators. Naturally one might wonder if a time symmetric choice of solutions is viable even at the classical level? Particularly, if one wishes to generate a theory in which charged particles interact with only the fields created by other charged particles (no self interaction), then the phenomena of radiation damping (emission) requires the existence of advanced fields arising from absorbers. If a source is accelerated, then a retarded (and advanced) field is created. The retarded field from the source interacts, at a later time, with the absorbers, which causes them to emit both retarded and advanced fields. The advanced fields from the absorbers, which are defined on the past light cone, interact with the source \emph{at the initial moment of emission}, and are such that if sufficiently many absorbers exist then the net advanced field at the position of the source exerts a force on the source particle so as to induce the expected amount of energy and momentum loss if the emission occurred via the standard picture \cite{WFI1} \cite{WFI2}.  

The transactional interpretation concerns the application of ``absorber theory" to quantum theory. Just as the equations of electromagnetism admits both retarded and advanced solutions, other relativistically invariant equations, in particular those with quantum mechanical significance (for example the Klein-Gordon equation), do so as well. An immediate difficulty presents itself -- the Schr\"odinger equation does not have advanced solutions. It is proposed that the transactional interpretation requires a fully relativistic theory. However, it is well known that the nonrelativistic limit includes both the Schr\"odinger equation \emph{and} its complex conjugate. The solutions to the former are purely retarded and the solutions of the later are purely advanced. Making use of the Schr\"odinger equation and its conjugate equation gives one access to a set of retarded and advanced solutions. An emitter is assumed to produce a retarded and advanced wave. Consider an emission process located at $(\mathbf{x}_1,t_1)$. The retarded wave from the emitter will be called $\psi_E (\mathbf{x},t>t_1)$ (verbally, the ``offer wave") and the advanced emitter wave will be called $\phi_E(\mathbf{x},t<t_1)$. The retarded wave $\psi_E$ interacts with an absorber at event $\mathbf{x_2},t_2$ which is also assumed to produce a retarded and advanced wave. We will refer to the retarded and advanced waves from the absorber as $\psi_A(\mathbf{x},t>t_2)$ (verbally, the ``confirmation wave") and $\phi_A(\mathbf{x},t<t_2)$. The advanced wave from the absorber propagates back to the source at the moment of emission. One assumes that the initial amplitude of the wave produced by the absorber is proportional to the incident amplitude, i.e.\begin{equation}\label{TI1}\phi_A(\mathbf{x},t) \propto \psi_E(\mathbf{x}_2,t_2)\upsilon_A(\mathbf{x},t)\end{equation} and $\upsilon_A$ is defined symmetrically with respect to emission so that $\upsilon_A(\mathbf{x},t)=\psi^*_E(\mathbf{x}+(\mathbf{x}_2-\mathbf{x}_1),t+(t_2-t_1))$. If these assumptions are met, then the amplitude of the advanced wave at the emission event is \begin{equation}\label{TIborn}\phi_A(\mathbf{x}_1,t_1)=|\psi_E(\mathbf{x}_2,t_2)|^2. \end{equation} It is in this manner that the Born rule arises in the transactional approach. The process of emission and re-emission (of an advanced wave) by the absorber is referred to as a ``transaction." It is claimed that an emitter continuously sends out offer waves and continuously ``hears" confirmation waves from absorbers. A transaction will only occur if $E=h\nu$ and other conservation laws are met \cite{Cramer1}\cite{Cramer2}\cite{Cramer3}.

There are severe deficiencies with this approach. Firstly, it should be clear that the transactional approach, which requires both retarded and advanced solutions, is best pursued in the context of a fully relativistic theory. However, the approach as described applies only to first quantized theories. It is not clear how to generalize this approach to second quantized field theories, if such a generalization is possible. To highlight this deficiency further, consider the assumption that the initial amplitude of the wave produced by the absorber is proportional to the incident amplitude, equation \eqref{TI1}. This assumption yields equation \eqref{TIborn}, which resembles the Born rule for nonrelativistic quantum mechanics. Does \eqref{TI1} hold for the Dirac equation? For the Dirac equation the appropriate product for constructing a probability distribution is \begin{equation}\rho_D(\mathbf{x},t)=\psi^\dagger(\mathbf{x},t) \psi(\mathbf{x},t)\end{equation} The advanced solutions of the Dirac equation are not, however, ``daggered." Therefore, if \ref{TI1} did hold for the Dirac equation then the resulting probability distribution would be wrong. A similar argument holds for the Klein-Gordon equation because there the appropriate probability is constructed as \begin{equation}\label{probKG}\rho_{KG}=\frac{i\hbar}{2m}(\psi^*\partial_t\psi-\psi \partial_t \psi^*),\end{equation}which is necessary for local conservation of (signed) probability. It is unclear how \eqref{probKG} could be arrived at from a transaction.

Absorber theory demands the existence of emitters (sources) and absorbers. Such objects exist in electromagnetism because generically the equations of electromagnetism are \emph{sourced}. Namely, the acceleration of charges sources electromagnetic radiation. However, equations with quantum mechanical interest are not sourced. Indeed, adding a source to the Schr\"odinger equation (or relativistically invariant equations) destroy local conservation of probability and therefore introduces non-unitary behavior. More bluntly, the mechanism of action by which emitters radiate and absorbers absorb radiation \emph{is included} within electromagnetism, but no such mechanism is included in quantum mechanics. Introducing ``emitters" and ``absorbers" whose action cannot be described within the theory is as problematic as the Copenhagen interpretation wherein a process (measurement) acts on a system (i.e. acts to collapse the wavefunction) in a manner that cannot be described within the theory (within the unitary sectory to be precise). One might ask about the nature of the emitters/absorbers. For example, with what properties must they be endowed so as to guarantee their ability to emit/absorb? We may say in a precise manner that charged matter can be made to source photons, but the transactional approach requires one to talk about sourcing particles or whole atoms or even whole molecules, as interference has been observed in each of these. 

How are we to interpret ``transaction?" It seems that the primary example of a transaction is a type of transition whereby a particle is emitted by an emitter and is found at a later time at the site of an absorber. The particle may then be re-emitted and the process will continue. Thusly, we are lead to conclude that transactions are stochastic processes, and consequently, that the transactional interpretation is more appropriately an example of an objective collapse theory wherein the collapses are induced by the presence of ``absorbers." Essential for our purpose is to note that such processes violate unitarity, which lead to experimental consequences. A related issue is the change in the wavefunction of a system that has been absorbed. In particular the transactional approach fails to specify if there is a change in the wavefunction upon absorption. This has a strong bearing on the ``repeatability of measurement" assumption of quantum mechanics under the Copenhagen interpretation, which states that a repeated measurement that occurs a short time after a first measurement will produce the same result. Without an assumption concerning the change in the wavefunction upon absorption, it seems that the transactional interpretation is incomplete. However, if such an assumption is added to the theory, then it is apparent that observation has a distinguished role, which is the measurement problem within the Copenhagen interpretation.   

It is claimed that a transaction occurs if $E=h\nu$ and other conservation laws are met. This statement is vacuous. The approach has not introduced a quantum equivalent for conservation laws. Namely, there is no mention of an operator formalism, nor is there a mention of an eigenstate-eigenvalue link. For example, if the wavefunction is a (near) plane wave, are we to interpret that as a system with well defined momentum? The statement that a transaction may occur if $E=h\nu$ (and other conservation laws are satisfied) suggests that the term ``transaction" refers to transition phenomena \emph{between} energy eigenstates. However, such transitions \emph{do not} conserve energy/momentum since the formalism does not include the electromagnetic field (i.e. the electromagnetic field functions as external perturbing agent). As a result, it is not clear if one should demand energy/momentum conservation or attempt to account for the energy/momentum loss to the electromagnetic field. There are two distinct possibilities and no clear answer as to which possibility one should appeal. As an example of an objective collapse theory, we see that, as expected, there is energy/momentum nonconservation at the moment of collapse (read transaction). That conservation laws must be \emph{added} as additional assumptions to this theory makes the transactional approach inelegant and implies incompleteness and experimental deficiency. Finally, we emphasize that without an operator formalism/eigenstate-eigenvalue link measurements of common physical quantities (energy, momentum, angular momentum, and others) are formally \emph{undefined} by the theory \cite{Cramer1} \cite{Cramer2} \cite{Cramer3}. 

We see that there are difficulties concerning the structure of the transactional interpretation. Most poignantly for this current work, however, is that it fails to resolve the measurement problem. Consider without loss of generality the transactional interpretation of the Schr\"odinger's cat paradox. The claim is made that the main difficulty with regard to this paradox is in answering \emph{when} collapse occurs. Since the transactional interpretation is (essentially) atemporal asking when a collapse occurs is an ill defined question. Instead, the transactional approach suggests that during the period when the cat is sequestered, a transaction may occur or may not, with probability provided by the Born rule, regardless of an act of observation (measurement). However, there are two dilemmas in this claim. Firstly, the formalism does not suggest that a transaction is ``atemporal." Indeed, the formalism suggests that transactions occur \emph{within well defined temporal intervals}. For example, explicit in the emission of a (half advanced half retarded) wave from an emitter is the time of emission. The absorption and emission of a confirmation wave also occur at well defined times so that it seems that the transaction occurs within the time interval, even if the precise time is undefined. Secondly, the unitary development of quantum mechanics predicts that the until the moment of wavefunction collapse the system is in a superposition and \emph{interference effects may be generated by measuring an appropriate variable}. The claim of the transactional approach stands in contradiction to standard quantum mechanics by denying the possibility of interference. Since superposition is an experimentally well verified facet, it is clear that there is a serious experimental shortcoming in the transactional interpretation. In this regard the transactional interpretation is an example of an objective collapse theory, and subsequently suffers the fwailings of such theories (see section \ref{OC}). Finally, it is incorrect to state that the paradoxical aspect of Schr\"odinger's cat results from attempting to answer when collapse occurs. The main difficulty is that certain physical processes (measurements) seem to induce collapse (definite properties) but these processes cannot be described from within the unitary sector of the theory. The transactional interpretation presents no description of the measurement process, and subsequently does not at all resolve the measurement problem.

In the current state of this formalism, one assumes that a single emitter interacts with a single absorber. Generically, however, one expects that this condition will not be met. Emitted waves should interact with many future absorbers and each of those absorbers should emit an advanced wave that interacts with many emitters ``in the past." That only one emission/absorption/confirmation process in considered is a serious shortcoming of this formalism that needs to be addressed by either (1) rigorous justification or (2) multiple emission/absoprtion/confirmation processes must be considered. 

\section{Time Symmetric Quantum Mechanics}
Time symmetric quantum mechanics is introduced in part to deal with a paradox concerning state vector collapse caused by simultaneous, but spatially separated measurements. A prime example of which is an experiment to test the Bell's inequalities. If the measurements occur simultaneously in an frame in which both observers (Alice and Bob, presumably) are at rest, then in comoving Lorentz frames the order of measurements may be seen to be reversed, raising the question as to which action collapsed the wavefunction. The interpretation is arrived at by first computing the probability for an intermediate measurement result $o_i$ at present time $t$ given that the initial state is pre-selected to be $\ket{\psi_{in}}$ and the final state is post-selected to be $\ket{\psi_{out}}$. Namely, 
\begin{eqnarray}P(o_i|\psi_{in}, t_{in}; \psi_{fin},t_{fin})= \nonumber \\ \frac{|\bra{\psi_{fin}}U^\dagger_{(t_{fin}\rightarrow t)} \ket{o_i}|^2|\bra{o_i}U_{(t_{in}\rightarrow t)}\ket{\psi_{in}}|^2}{\sum_j\left(|\bra{\psi_{fin}}U^\dagger_{(t_{fin}\rightarrow t)} \ket{o_j}|^2|\bra{o_j}U_{(t_{in}\rightarrow t)}\ket{\psi_{in}}|^2\right)}\label{TS1} \end{eqnarray} The formalism may be further refined by the use of \emph{two states}, which are operators of the form \begin{equation}\rho = \ket{\psi_1(t)}\bra{\psi_2(t)},\end{equation} or linear combinations thereof \cite{A1} \cite{A2} \cite{A3} \cite{A4}. 
Whether equation \eqref{TS1} is useful in understanding quantum mechanics is not of concern for this work. We ask, instead, whether it may form the basis of an interpretation that resolves the measurement problem and other related difficulties. It is apparent that \eqref{TS1} makes use of the Born rule, without an attempt at derivation or explanation. We are only lead to assume that an interpretation based on \eqref{TS1} must also assume the Born rule, and as a result, does not provide any insight into measurement as a physical process nor does it obviate the need for collapse-like (more generally nonunitary) behavior. While it may be objected that the possibility of deriving the Born rule from the overlap of retarded and advanced waves, in great similarity to the transactional interpretation, still exists, pursuit of such a ``derivation" raises further issues (see Section \ref{TI}). Equation \eqref{TS1} still indicates that a nonunitary transformation of the wavefunction occurs within the interval $[t_i, t_f]$. What spurs such an evolution? The formalism does not dictate the mechanism. If, as one might naturally assume, they occur spontaneously, then the result is an objective collapse theory, which could easily be ruled out experimentally (see section \ref{OC}). We emphasize, in passing, that the measurement problem does not appear to stem from a problem with the treatment of time so that variations of standard quantum mechanics with different approaches to time will not lead to a resolution of the measurement problem. 

\section{Concluding Remarks}

We challenge the viability of common interpretations of quantum mechanics. We have argued, and where possible, demonstrated, that each interpretation exhibits significant deficiency, in the form of fine-tuning problems/ad hoc assumptions, internal inconsistencies, incompleteness, disagreement with experiment, among others. We raise an obvious question, but do not attempt an answer here. Is there an interpretation (more appropriately reinterpretation) \emph{or} minor modification of quantum mechanics that will maintain full agreement with experiment (both current and future) and will resolve the measurement problem without introducing new issues?

\bibliographystyle{unsrtnat}

\end{document}